\documentclass[11pt,a4paper]{article}
\usepackage{amsmath}
\usepackage{dsfont}
\usepackage{bm}
\usepackage{esint}
\usepackage{subcaption}
\usepackage{accents}
\usepackage{mathrsfs}
\usepackage{amsbsy}

\usepackage{a4wide,graphicx,times,psfrag,wrapfig,sidecap}
\usepackage{cite}
\usepackage[colorlinks=true,linkcolor=black, citecolor=black,
urlcolor=black]{hyperref}
\numberwithin{equation}{section}
\makeatletter \let\old@startsection=\@startsection
\renewcommand{\@startsection}[6]
{\old@startsection{#1}{#2}{#3}{#4}{#5}{#6\mathversion{bold}}}
\makeatother

\def\pre{Phys. Rev. E}
\def\prl{Phys. Rev. Lett.}
\def\physrep{Phys. Rep.}
\def\<{\langle}
\def\>{\rangle}

\def\Im{{\rm Im}}
\def\Re{{\rm Re}}

\newcommand\encadremath[1]{\vbox{\hrule\hbox{\vrule\kern8pt
\vbox{\kern8pt \hbox{$\displaystyle #1$}\kern8pt}
\kern8pt\vrule}\hrule}} \def\enca#1{\vbox{\hrule\hbox{
\vrule\kern8pt\vbox{\kern8pt \hbox{$\displaystyle #1$} \kern8pt}
\kern8pt\vrule}\hrule}}
  \usepackage{bm}

\def\XXint#1#2#3{{\setbox0=\hbox{$#1{#2#3}{\int}$}
     \vcenter{\hbox{$#2#3$}}\kern-.5\wd0}}

\begin{document}

\begin{center}
{\Large  Whitham Approach to Certain Large Fluctuation Problems in Statistical Mechanics  }

\vspace{10mm}

Eldad Bettelheim  \\[7mm]

Racah Institute of Physics, Hebrew University of Jerusalem, Edmund J Safta Campus\\ 91904 Jerusalem, Israel\\ 

\vspace{20mm}

\end{center}

\vskip9mm
\begin{abstract}
We show the relationship between the strongly non-linear limit (also termed the dispersionless or the Whitham limit) of the macroscopic fluctuation theory of certain statistical models and  the inverse scattering method. We show that in the strongly non-linear  limit the inverse scattering problem  can be solved using the steepest descent method of the associated Riemann--Hilbert problem. The importance of establishing this connection, is that the  equations in the strongly non-linear limit can often be solved exactly by simple means, the connection then provides a limit in which one can solve the inverse scattering problem, thus aiding potentially the exact solution of a particular large deviation problem.     
\end{abstract}
 
\section{Introduction}
In this paper we study some mathematical aspects of large deviation problems that have garnered some interest in recent years. Such problems include the Kardar-Parisi-Zhang problem\cite{KPZ}, the Kipnis-Marchioro-Presutti model\cite{KMP}, the symmetric exclusion process\cite{Derrida:SSEP}, and more.  Particularly interesting in this regard is the availability of exact solutions \cite{Derrida:NonEq:Flucutations:Larege:Deviation}. Even in the absence of such exact solution one may use macroscopic fluctuation theory \cite{Bertini:DeSole:Gabrielli:MFT} to study the question of large deviation.  One such case is the Kardar-Parisi-Zhang problem in which the macroscopic fluctuation theory leads to the following equations\cite{Janas:Meerson:Kamenev}: 
\begin{align}
&\label{NLS1}\partial_t q=\partial_x^2 q+\frac{1}{4}q^2p\\
&\label{NLS2}\partial_t p=-\partial_x^2 p-\frac{1}{4}qp^2
\end{align}
where the boundary value to be solved is given by
\begin{align}
p(x,1)=\Lambda\delta(x), \quad q(x,0)=f(x).\label{BoundaryConditions}
\end{align}
for some given $f(x)$. Typically $f(x)$ is chosen to be either constant (flat initial conditions) or itself proportional to a delta function, but other boundary conditions may be considered as well.
Here
$f(x)$ has the meaning of the initial shape of the Kardar-Parisi-Zhang interface and the boundary condition for $p(x,1)$ allows to compute the appropriate generating function for the interface at final time $t=1$. The variable $\Lambda$ serves as the parameter of the generating function\cite{Janas:Meerson:Kamenev}.  

Due to the similarity of the formulation of the large deviation problem across different models, we shall concentrate here on the example where the equations to be solved are given by Eqs. (\ref{NLS1}, \ref{NLS2}) and the boundary conditions are given in Eq. (\ref{BoundaryConditions}). Other problems have a very close formulation, either directly \cite{Bettelheim:Meerson:JStatMech,Bettelheim:Meerson:PRL,Doussal:Krajenbrink:CompetingPaper} or through non-trivial manipulation \cite{Mallick:Moriya:Sasamoto}. For example, for the Kipnis-Marchioro-Presutti  model one has to consider rather the derivative nonlinear Sch\"odinger equation, a variant of Eqs. (\ref{NLS1}, \ref{NLS2}). Nevertheless, in the formal sense the Eqs. (\ref{NLS1}, \ref{NLS2}) and the derivative non-linear Schr\"odinger equations have so much in common, that  the methods provided here are easily modified to deal with that problem as well. As a result, we prefer, for the sake of brevity, to describe here the method for the particular case of the nonlinear Schr\"odinger equation, Eqs. (\ref{NLS1}, \ref{NLS2}) , without spelling out how the method must be generalized to other cases.  

To solve the boundary value problem described above,  one may apply the inverse scattering method\cite{Krajenbrink:Doussal:Flat:Brownian,Krajenbrink:Doussal:InverseScattering,Mallick:Moriya:Sasamoto,Bettelheim:Meerson:JStatMech,Bettelheim:Meerson:PRL,Doussal:Krajenbrink:CompetingPaper}, which we will describe below. Despite being integrable, or "completely soluble", there is no recipe to solve a particular boundary problem for the nonlinear Sch\"odinger equation,  such as the problem presented in Eq. (\ref{BoundaryConditions})\ above, and it is often necessary to rely on luck to find such a solution. 

Nevertheless, certain limits are tractable. For example, if the fields $p$ and $q$ are small, such that the non-linear terms may be neglected, then a solution may be easily found. Another tractable limit, is the limit where the fields are large, such that the non-linear term dominates over the dispersive term in the non-linear Schr\"odinger equation, leading to non-viscid equations. In such limits it may be instructive to obtain a full solution to the large deviation problem, provided one knows how to perform the scattering transform to the solution obtained in the special limits. Then one may examine the scattering data, in order to have an educated guess for the full solution. 

The current paper concerns itself with this program, for the case of the strongly non-linear limit (the linear limit is much easier to deal with, and is usually also much less instructive in finding exact solution, despite its physical importance). In certain cases in this limit, it is possible simply to drop the dispersive terms in the equations, and thus deal with the inviscid equations that arise. This case is most instructive, although the method we present is not restricted to this case. In fact, often solitons or oscillatory features may appear in the strongly non-linear limit, and the method described below is suitable to deal with this situation as well. Despite this, even in the case of sharp or oscillatory features, the solution may still feature regions in space-time where the inviscid equations hold with no dispersive terms, such that the inviscid equations contain a large amount of information even in the presence of sharp or oscillatory regions .This situation is well known in the strongly  non-linear limit, which is often also called  the "dispersionless limit", or the "Whitham limit"\cite{Whitham:1966:NDW}.

We proceed  to solve the inverse scattering problem in the strongly non-linear limit. This is done by applying the map from  inverse scattering problem to the Riemann-Hilbert problem, and then solving the Riemann-Hilbert problem thus obtained using the steepest descent method following closely  Ref. \cite{Deift:Zhou:Steepest:Descent}. 

The result of the work is then a relation between the strongly non-linear equations and the inverse scattering problem. In addition to the potential utility of this approach in discovering new solutions, the connection thus obtained also has additional advantages. Among them we list two here. First, the method allows to find a systematic expansion in a small parameter around the strongly non-linear solution. This is not done in the current paper, since to apply this approach requires to introduce even more tools associated with Riemann-Hilbert problem, and we leave that for future work. Secondly, the method allows to find solutions which feature oscillatory regimes for  the fields $p$ and $q.$ Admittedly, though, the physical meaning of such solutions is not known at present, beyond their importance for the case of periodic boundary conditions\cite{Smith:Sasorov:Meerson:Periodic,Zarfaty:Meerson:Periodic}.

To demonstrate our approach we work out the  invscid flows in the case of the Kardar-Parisi-Zhang problem with flat initial conditions as worked out in Ref. \cite{Meerson:Katzav:Vilenkin}, and show how these are connected to the exact solution given in Ref. \cite{Krajenbrink:Doussal:Flat:Brownian}.

\section{Na\"{i}ve Inviscid Limit}

We first study the inviscid limit of Eqs. (\ref{NLS1}, \ref{NLS2}).
Since these are a variant of the nonlinear Schr\"odinger equations (where $p$ and $q$ are now not complex conjugates of each other but rather real) it is natural to consider the Madelung transformation to the fields such as to obtain hydrodynamic type equations for appropriately defined density and velocity fields, familiar from quantum mechanics. In this case it is also appropriate to call such a transformation a  Cole-Hopf transformation. We apply the following\cite{Janas:Meerson:Kamenev}: 
\begin{align}
\label{ColeHopf}q= e^{-\frac{1}{2}\int^x v},\quad p=-2 \rho e^{\frac{1}{2}\int^x v}
\end{align}
 to obtain:
\begin{align}
\partial_t \rho+\partial_x(\rho v)&=-\partial_x^2 \rho\\
\partial_t v+v\partial_x v-\partial_x \rho&=\partial^2_xv.
\end{align}
One may now discard the terms on the right hand side of the equation, which are of higher derivative, and thus may be termed, dispersive, or viscous terms (thus obtaining the na\"ive inviscid limit). The resulting equations are those of a fluid with density $\rho$ and velocity $v$ and pressure given by $\frac{-\rho^2}{2}$ \cite{Meerson:Sasorov:2014}:
\begin{align}
\label{inviscid1}\partial_t \rho+\partial_x(\rho v)&=0,\\
\label{inviscid2}\partial_t v+v\partial_x v-\partial_x \rho&=0.
\end{align}
These equations may be subjected to another transformation to bring them into the form of Riemann invariants. The Riemann invariants, $\lambda_1$, $\lambda_2$ are defined as follows:
\begin{align}
\lambda_i=\frac{\imath v}{2} \mp\sqrt{\rho},\label{transformToInvariants}
\end{align}
whereby the equations that result by this transformation are given by:\begin{align}
&\label{RiemannInvariant1}\partial_t \lambda_1=\left(\frac{3}{2}\lambda_1+\frac{1}{2}\lambda_2\right)\imath \partial_x\lambda_1\\
&\label{RiemannInvariant2}\partial_t \lambda_2=\left(\frac{3}{2}\lambda_2+\frac{1}{2}\lambda_1\right)\imath\partial_x\lambda_2.
\end{align}
These equations are typical of Dispersionless integrable equations and are in fact the Whitham\cite{Whitham:1966:NDW} universal equations written through Riemann invariants\cite{83:Krichever:Averaging,88:Flaschka:KdV:Avging}. 
\section{The Inverse Scattering Method for Large Fluctuation Problems}
Now we turn our attention to the inverse scattering method, in order to understand how the inviscid solutions translate in terms of the scattering data. 

The inverse scattering method relies on the fact that by introducing an auxiliary parameter $k$ and defining two $k,$ $x$ and $t$ dependent matrices \cite{Faddeev:Book:Hamiltonian:Methods}
\begin{align}
U(k)=\begin{pmatrix}- \frac{\imath k}{2}   & -\frac{p}{8}  \\
q  & \frac{\imath k}{2} \\
\end{pmatrix}, \quad V(k)=\begin{pmatrix}\frac{k^2}{2}-\frac{pq}{8} & \frac{p'-\imath kp}{8} \\
p'+\imath kp  & -\frac{k^2}{2}+\frac{pq}{8} \\
\end{pmatrix},
\end{align}
one may recast the nonlinear Schr\"odinger equations, Eqs. (\ref{NLS1},\ref{NLS2}),
 in the following form:
\begin{align}
\partial_t U-\partial_x V+[U,V]=0.\label{Consistency}
\end{align}
Indeed if these equations are to hold for any $k,$ $x$ and $t$ then Eqs. (\ref{NLS1},\ref{NLS2}) must be satisfied. Eq. (\ref{Consistency}) has the form of a consistency condition for the following equations for a $2\times2$ matrix, $P$:
\begin{align}
\partial_x P=UP, \quad \partial_t P=VP.\label{Peqs}
\end{align}
Indeed, if Eq. (\ref{Consistency})
is satisfied then a solution for  $P$ exists.  

If one assumes the $p$ and $q$ tend to $0$ at $x\to\pm\infty$ for any $t$, then $P$ has solutions in that region in the form of plane waves $P\sim e^{-\sigma_z\frac{\imath kx}{2}}T$, for any $x-$independent matrix $T$. This leads naturally to consider the scattering problem. The problem reads as follows: given $k$ and a plane wave solution at $x\to -\infty$ of the form $P\sim e^{-\sigma_z\frac{\imath kx}{2}}$, find $T(k,t)$ as featured in the asymptotic behavior  of $P$\ at $x\to +\infty$ as follows $P\sim e^{-\sigma_z\frac{\imath kx}{2}}T(k,t)$. 

Writing
\begin{align}
T(k,t)=\begin{pmatrix}a_+(k,t) & \tilde b(k,t) \\
b(k,t) & a_-(k,t) \\
\end{pmatrix},
\end{align}
one can easily find the time dependence of the elements as follows: $a_\pm(k,t)=a_\pm (k,0),$  $b(k,t)=e^{-k^2 t}b(k,0)$ and  $\tilde b(k,t)=e^{k^2 t}\tilde b(k,0).$ This can be shown by advancing the plane wave solutions in time at $x=\pm\infty$. One obtains that at large $|x|$ the solution behaves as  $P\sim e^{\sigma_z\left(\frac{k^2t}{2}-\frac{\imath kx}{2}\right)},$ namely advancing the plane waves in time amounts to applying the matrix $ e^{\sigma_z\frac{k^2t}{2}} .$ To obtain then the scattering matrix $T(k,t)$ at any finite time $t$ it is enough to know the scattering matrix at time $t=0$. Indeed one may deduce the scattering at time $t$ by  first considering the plane waves at $x=-\infty$ at time $t$ and then  rewinding these plane waves from time $t$ to time $0$\ by applying the matrix $e^{-\sigma_z\frac{k^2t}{2}},$ then one lets the waves scatter from $-\infty$ to $\infty$ by applying the matrix $T(k,0)$ and finally one advances the plane waves from time $t=0$ back to time $t$\ by applying the matrix $e^{\sigma_z\frac{k^2}{t}}$ . The application of the matrices is always from the left such that we get  $T(k,t)=e^{\sigma_z\frac{k^2t}{2}}T(k,0)e^{-\sigma_z\frac{k^2t}{2}}$.
Furthermore, it can be shown that  $\det T(k)=a_+a_-+b\tilde b=1$ and $a_\pm$ can be shown to be analytic in the upper and lower half planes, respectively.  
\section{The Riemann-Hilbert Approach  to Inverse Scattering}
Let us find a matrix solution  $P(x,y)$ to Eq. (\ref{Peqs}) with boundary conditions:
\begin{align}
\partial_x P(x,y;k)=U(k)P(x,y;k), \quad P(x,x;k)=\begin{pmatrix}e^{\frac{\imath k x}{2}} &  \\
 & e^{-\frac{\imath k x}{2}} \\
\end{pmatrix}
\end{align}

This function has the property \begin{align}P(x,y;k)P(y,z;k)=P(x,z;k)\label{GroupProperty}.\end{align} This allows one to define:
\begin{align}
&T(k)\equiv \lim_{y,-x\to\infty} \begin{pmatrix}e^{-\frac{\imath k x}{2}} &  \\
 & e^{\frac{\imath k x}{2}} \\
\end{pmatrix}P(x,y;k)\begin{pmatrix}e^{-\frac{\imath k y}{2}} &  \\
 & e^{\frac{\imath k x}{2}} \\
\end{pmatrix},\\
&P^+(x;k)=\lim_{y\to\infty}P(x,y;k)\begin{pmatrix}e^{-\frac{\imath ky}{2}} &  \\
 & e^{\frac{\imath k y}{2}} \\
\end{pmatrix}\label{Pplus},\\&P^-(x;k)=\lim_{y\to-\infty}P(x,y;k)\begin{pmatrix}e^{-\frac{\imath ky}{2}} &  \\
 & e^{\frac{\imath k y}{2}} \\
\end{pmatrix}.\label{Pminus}
\end{align}
     Using the property in Eq. (\ref{GroupProperty}) one obtains:
\begin{align}
P^+(x;k)=P^-(x;k)T(k).
\end{align}
It is then possible to take this equation and rearrange the elements\cite{Faddeev:Book:Hamiltonian:Methods} such as to write the following:
\begin{align}
G_-=G_+ G,
\end{align}
where the different objects are defined as follows:
\begin{align}
&G=\begin{pmatrix}1 &\frac{ \tilde b}{a_-} \\
\frac{b}{a_+}  & \frac{1}{a_+a_-} \\
\end{pmatrix}\label{},\quad G_+=\begin{pmatrix}\frac{P^-_{11}}{a_+} & P^+_{12} \\
\frac{P^-_{21}}{a_+} & P^+_{22} \\
\end{pmatrix}, \quad G_-=\begin{pmatrix}P^+_{11} & \frac{P^-_{12}}{a_-} \\
P^+_{21} & \frac{P^-_{22}}{a_-} \\
\end{pmatrix},\label{GandGpGm}
\end{align}
and we have dropped denoting the dependence on $k$ of $a_\pm$ and $b,\tilde b$  for brevity and it is implicitly assumed  that these objects are to be evaluated at $t=0$. 
The matrices $G_+$ and $G_-$ can then be shown to be analytic in the upper half and lower half planes, respectively. Such that we obtain the following Riemann-Hilbert problem:

\begin{itemize}
\item{$G_\pm$ are analytic in the upper and lower half planes, respectively.}
\item{On the real axis $G_-=G_+ G$.}
\item{$G_\pm(z)\to \mathds{1}+O(1/z)$ as $z\to\infty$}
\end{itemize}

Given $G$, if one is able to solve the Riemann-Hilbert problem, then one obtains immediately the fields $p$ and $q.$ Indeed, $G_\pm$ satisfy:
\begin{align}
\partial_xG_\pm G_\pm^{-1}=\begin{pmatrix}-\frac{\imath k}{2}  & -p \\
\frac{q}{8} & \frac{\imath k}{2} \\
\end{pmatrix} \label{Gpmxderiv},
\end{align} 
such that $p$ and $q$ are easily extracted from this equation.  However, the Riemann-Hilbert problem is often intractable. Nevertheless, it has asymptotes of the solution may be found by using the method of steepest descent presented in the following.  
\subsection{The Riemann-Hilbert Problem in Macroscopic Fluctuation Theory }
Let us first study some particular properties of the Riemann-Hilbert problem in the case of macroscopic fluctuation theory. In this case $p$ is proportional to a delta function, so we write 
$p(x,1)=-8\Lambda\delta(x).$ 
\begin{align}
\partial_xP=\begin{pmatrix}-\frac{\imath k}{2}  & \Lambda \delta(x) \\
q & \frac{\imath k}{2} \\
\end{pmatrix} P.\label{PwithDelta}
\end{align}
This is easily solved, for example, by:
\begin{align}
P^-=\begin{cases}\begin{pmatrix}e^{-\imath\frac{  kx}{2}} & 0 \\
e^{\imath\frac{  kx}{2}}I(-\infty,x) & e^{\imath\frac{  kx}{2}} \\
\end{pmatrix} & x<0 \\
\begin{pmatrix}e^{-\imath\frac{  kx}{2}}\left(1+\Lambda I(-\infty,0))\right) & \Lambda e^{-\imath\frac{  kx}{2}} \\
e^{\imath\frac{  kx}{2}}\left(I(-\infty,x)+\Lambda I(-\infty,0)I(0,x)\right) & e^{\imath\frac{  kx}{2}}\left(1+\Lambda I(0,x))\right) \\
\end{pmatrix} & x>0 \\
\end{cases}
\end{align}
where $I(a,b)=\int_a^b e^{-\imath k x'} q(x')dx'$. So that we get:
\begin{align}
T(k,1)=\begin{pmatrix}a_+(k) & \tilde b(k,1) \\
b(k,1) & a_-(k) \\
\end{pmatrix}=\begin{pmatrix}1+ \Lambda I(-\infty,0) & \Lambda \\
I(-\infty,\infty)+ \Lambda I(-\infty,0) I(0,\infty) & 1+ \Lambda I(0,\infty) \\
\end{pmatrix}
\end{align}
where we have thus obtained:
\begin{align}
a_+(k,t)=1+ \Lambda I(-\infty,0), \quad a_-(k,t)=1+ \Lambda I(0,\infty)\quad \tilde b(k,1)=\Lambda
\end{align}
 Thus
given the time dependence of the elements of $T$ discussed above, we have:\begin{align}
T(k,t)=\begin{pmatrix}a_+(k) & \Lambda e^{-\imath kx +k^2(t-1)} \\
\frac{ e^{\imath kx-(t-1)k^2}(a_+(k)a_-(k)-1)}{ \Lambda   } & a_-(k) \\
\end{pmatrix}
\end{align}

We find $G,$ the jump matrix associated with the Riemann-Hilbert problem, according to Eq. (\ref{GandGpGm})\begin{align}
&G=\begin{pmatrix}1 & \Lambda e^{-\imath kx+(t-1)k^2-\log(a_-) } \\
-\frac{ e^{\imath kx-(t-1)k^2+\log(a_-)}}{ \Lambda   }\left(1-\frac{1}{a_+a_-}\right)  & \frac{1}{a_+a_-} \\
\end{pmatrix}
\end{align}

At $t=1$ one can find $G_\pm$ explicitly through the Fourier transform of $q$. This is done by solving Eq. (\ref{PwithDelta}) and inserting that solution into Eqs. (\ref{Pplus},\ref{Pminus},\ref{GandGpGm}). This gives the following expressions:

\begin{align}
& \label{Gpexact}G_+=\begin{cases}\begin{pmatrix}1 & 0 \\
\frac{ e^{\imath k x}}{ \Lambda   } \left(1+\Lambda I(0,x)-a^{-1}_+  )\right)  & 1 \\
\end{pmatrix} & x>0\\
\begin{pmatrix}\frac{1}{a_+} & -\Lambda e^{-\imath kx } \\
\frac{e^{\imath kx}I(-\infty,x)}{a_+} & 1-\Lambda I(0,x) \\
\end{pmatrix} & x<0   
\end{cases}\\
&\label{Gmexact} G_{-}=\begin{cases}\begin{pmatrix}1 & \frac{\Lambda e^{-\imath kx }}{a_-} \\
 -e^{\imath k x}  I(x,\infty)  &  \frac{  1+\Lambda  I(0,x)}{  a_-   }   \\
\end{pmatrix} & x>0\\
\begin{pmatrix}a_- & 0 \\
e^{\imath kx}  \left(\Lambda I(0,\infty)I(0,x)-I(x,\infty)\right)  & \frac{1}{a_-} \\
\end{pmatrix} & x<0   
\end{cases}
\end{align}
It is then easy to ascertain that the matrices $G_\pm$ are indeed analytic in the upper and lower half planes respectively and that $G_-=G_+G$.

\section{The Steepest Descent Method}

We now want to derive the strongly non-linear limit of the inverse scattering problem, this limit is in fact the one considered in Ref. \cite{Deift:Zhou:Steepest:Descent} where the steepest descent method to the Riemann-Hilbert problem may be used, but before going on to describe the steepest descent method for the Riemann-Hilbert problem relevant for all time $t$, we concentrate on its form in the case $t=1$. At this time-point, the Riemann-Hilbert problem is solved making use of Fourier transforms of $q(x)$ and its restriction to subintervals of $\mathds{R }, $ see Eqs. (\ref{Gpexact}, \ref{Gmexact}). Such that the steepest descent method here is really nothing but the usual steepest descent method for the Fourier integral, which is intimately related to the Legendre transform.   

A key to finding the solution in the inverse scattering method is to find $b(k,t)$ from which all other elements of the scattering matrix $T$\ can be found. Indeed, since we know $\tilde b(k,t)=\Lambda e^{k^2(t-1)}$ and due to the unimodularity of the scattering matrix, which results in the equation $a_+a_--b\tilde b=1$, one may easily find $a_+a_-$ given $b(k,t)$. Then using the fact that $a_\pm$ are analytic in the upper and lower half $k-$planes respectively, one can write:
\begin{align} 
a_\pm(k)=\mp\int \frac{\log\left(1+\Lambda b(\mu,1)\right)}{k\pm\imath 0^+-\mu} \frac{d\mu}{2\pi\imath} 
\end{align} 

Now we assume 
\begin{align}
\log(\Lambda b(k,1))=m(k)\label{mExactDef}
\end{align}
and that $|m(k)|$ is large. Let $S=\{k\in \mathds{R}|\Re (m(k))>0\}$ . We assume that $S$\ is a union of a finite number of intervals. We also have the condition $m(\mu)=m^*(-\mu)$, which is required to obtain a real solution.

Then we may write approximately
\begin{align}
\log a_\pm=\mp\int_S \frac{m(\mu)}{k\pm\imath 0^+-\mu}\frac{d\mu}{2\pi\imath }.
\end{align}
At $t=1,$ the solution of the Riemann-Hilbert problem is solved by the Fourier transform of $q(x),$ which is denoted by $I(-\infty,\infty)$. And the steepest descent method yields nothing but the saddle point equations for this integral. Let us assume $k$ is in the lower half plane, since $a_-$\ tends to one in this plane then $a_+$ must be large  we may conclude:
\begin{align}
a_+(k)=1+\Lambda I(-\infty,0)\sim\Lambda I(-\infty,0)\sim\Lambda I(-\infty,\infty).  
\end{align}
So that $q(x)$ is the inverse Fourier transform of $a_+$ and we find the saddle point equations:
\begin{align}
 x=-\imath \log' a_\pm(k^*(x))\label{teq1GeneralEq}=\pm\int_S \frac{m'(\mu)}{k^*\pm\imath 0^+-\mu}\frac{d\mu}{2\pi }.
\end{align}
Solving this equation for $k$ one may then find $q(x)$ by substituting 
\begin{align}
q=\exp{\imath\left[ k^*(x)x-\imath\log(a_\pm(k^*(x))\right]}.
\end{align}
The term in the square brackets is of course the Legendre transform of the $\imath \log a_\pm$, which we may denote by $\Phi(x)$. Of course $\Phi'(x)=k^*(x)$ so that we can also write $q=\exp \imath \int^xk^*(x)$. Making the identification $k^*(x)=\imath \frac{v(x)}{2}$ one obtains one half of the Cole-Hopf transformation, Eq. (\ref{ColeHopf}), $q(x)=e^{-\frac{1}{2}\int^xv}.$ Since $\rho=0$ at time $t=1$ for all $x\neq0$ then  $p(x)=0$ and one cannot identify the other half of the Cole-Hopf transformation. At this point the identification  $k^*=\imath\frac{v}{2}$ is merely suggestive, but later we shall be able to make this identification on more general grounds.

This rather trivial steepest descent approach for $t=1$ does not rely on the Riemann-Hilbert problem at all, since for this specific time $\tilde b(k,1)=\Lambda $ is a constant and the problem becomes essentially linear. Nevertheless, during the full time evolution The Riemann-Hilbert problem becomes more complex and requires a more involved solution which we discuss in the next sub-section.  

\subsection{Reduction of the Problem}

The steepest descent method of the Riemann-Hilbert problem includes first deforming the contour where the matrix $G_\pm$ have a jump discontinuity (the "jump contour", or the "Riemann-Hilbert contour"), such as to simplify the Riemann-Hilbert problem greatly. This in analogy to the steepest descent method for integrals, but here this step includes some matrix manipulation and a doubling or tripling of the jump contour, unlike the simpler steepest descent methods for integrals. After this is achieved the simpler Riemann-Hilbert problem is solved. This sub-section concerns itself with the first step, namely the reduction of the Riemann-Hilbert problem to a simpler one. It is already in this step that we will be able to identify the Riemann invariants $\lambda_1,$ $\lambda_2,$ or equivalently, the fields $\rho$ and $v$, but only in the inviscid case. In the more general case, the identification of the fields is more complicated and may only be deduced after performing the second step of actually solving the reduced Riemann-Hilbert problem.

The steepest descent method can be applied when have $\log (a_+a_-)$ is large. In this case the matrix $G$ takes the approximate form:
\begin{align}
G\simeq\begin{pmatrix}1 &\Lambda e^{-\imath kx + k^2 (t-1)-\log(a_-)} \\
-\frac{ e^{\imath kx - k^2 (t-1)+\log(a_-)} }{\Lambda} & \frac{1}{a_+a_-} \\
\end{pmatrix} 
\end{align}

\begin{figure}[ht]
\begin{center}
\includegraphics[width=0.60\textwidth,clip=]{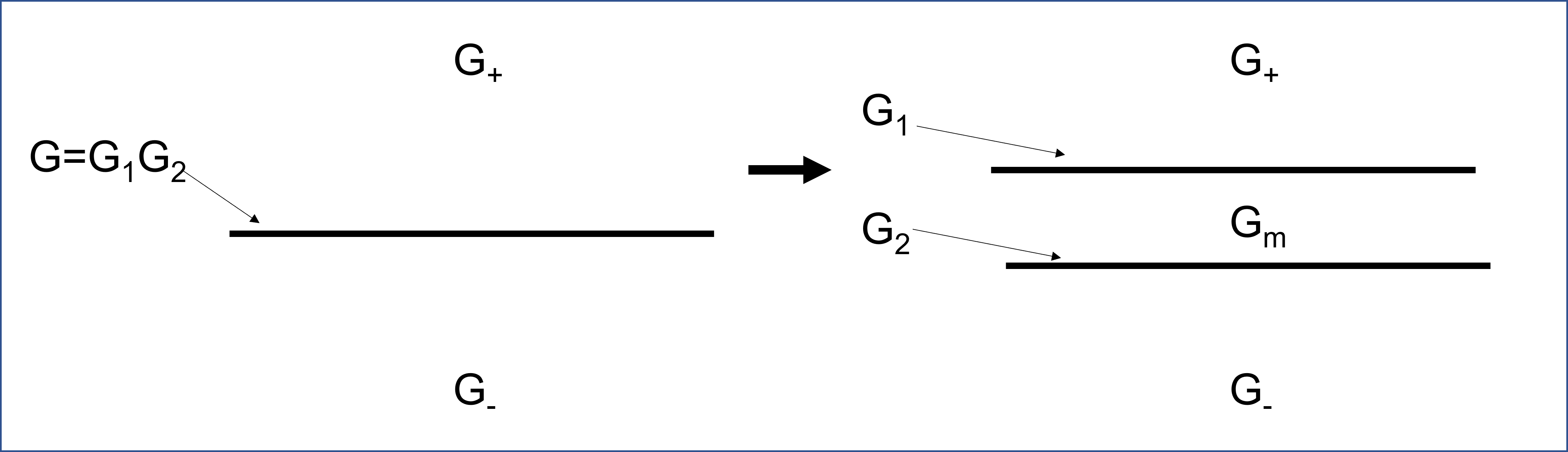}
\end{center}
\caption{We have dropped the tildes in this figure. On the left the real axis is denoted by a heavy line and the matrices $G_+$ and $G_-$ are shown above and below it respectively. The jump matrix is $G$ and it can be decomposed into $G_1G_2. $ The jump can then be separated into contours each one with its own jump matrix, $G_1$ and $G_2$. The jump then proceeds in two steps $G_+$ jump over to $G_m$ on the upper contour and $G_m$ jumps over to $G_-$ on the lower one. However, since the matrices $G_1$ and $G_2$ tend to the identity, the jump in fact disappears and $G_+$ becomes smoothly $G_-$\ as the real axis is crossed .   }
\label{RHdoubling}
\end{figure}

 Now we follow the classic method of Ref. \cite{Deift:Zhou:Steepest:Descent} to solve this problem approximately. The first step is to note that if we find two functions, $g_\pm$ analytic in the upper and lower half planes, respectively, and tending to $1$ at infinity at both half-planes,  then the following transformation of $G_\pm$ and $G$ leads to a new Riemann-Hilbert problem which has the same formulation as the original one. The transformation reads:\begin{align}
&\tilde G= \begin{pmatrix}e^{g_{+}} &  \\
 & e^{-g_{+}} \\
\end{pmatrix}G \begin{pmatrix}e^{-g_{-}} &  \\
 & e^{g_{-}} \\
\end{pmatrix}=\\&=\begin{pmatrix}e^{g_{+}-g_{-}} &  \Lambda e^{-\imath kx-\log(a_-)+g_{+}+g_{-}} \\
-\frac{  1}{ \Lambda   }e^{\imath k x+\log(a_-)-g_{+}-g_{-} } & e^{g_{-}-g_{+}-2\log\Lambda+k^2} \\
\end{pmatrix}\\
&\tilde G_+=G_+ \begin{pmatrix}e^{-g_{+}} &  \\
 & e^{g_{+}} \\
\end{pmatrix}, \quad \tilde G_-= G_- \begin{pmatrix}e^{-g_{-}} &  \\
 & e^{g_{-}} \\
\end{pmatrix}.
\end{align}
Let us denote:
\begin{align}
\imath h(k)=\imath k x-\log(a_-(k))-g_{+}(k)-g_{-}(k)-\log(\Lambda).
\end{align}
One can search for such functions $g_\pm$ such  that on some part of the jump contour, $g_+-g_-=0$ and that there  $h'(k)>0$ on the contour where the jump occurs. In this case, one may make use of the following decomposition of $\tilde G$:
\begin{align}
&\tilde G=\begin{pmatrix}1 &   e^{-\imath h(k)} \\
-e^{\imath h(k) } & \frac{1}{a_+a_-} \\
\end{pmatrix}\simeq G_1G_2\\&\tilde G_1=\begin{pmatrix}1 & 0 \\
-e^{\imath h(k) } & 1 \\
\end{pmatrix}, \quad \tilde G_2=\begin{pmatrix}1 & e^{-\imath h(k)} \\
0 & 1 \\
\end{pmatrix},
\end{align}
where one have used that $a_+a_-$ is large. One may then separate the jump contour into two contour, one in which the matrix $\tilde G_+$ jumps by $\tilde G_1$ and the other where the matrix $\tilde G_+$ jumps by $\tilde G_2$ . As a second step we deform each of the contours, that of $\tilde G_1$ towards the upper-half plane and that of $\tilde G_2$ towards the lower  half plane. The condition $ h'(k)>0$ then ensures that the off-diagonal terms in both $\tilde G_1$ and $\tilde G_2$ tends quickly to zero, and as such the jump matrix becomes the identity matrix, namely there is no jump. The procedure is illustrated in Fig.\ref{RHdoubling}. 
 Thus the part of the jump contour in which we were successful to find  $g_+-g_-=0$ and $h'(k)>0$ simply disappears. 

Secondly, with the same $g_\pm$ there may be a region where $g_--g_+=\log(a_+a_-)$ and $h'(k)<0,$ allowing us to write $\tilde G$:
\begin{align}
&\tilde G\simeq \begin{pmatrix}0 &   e^{-\imath h(k)} \\
-e^{\imath h(k) } & 1 \\
\end{pmatrix}=\begin{pmatrix}1 & e^{-\imath h(k) } \\
0 & 1 \\
\end{pmatrix}\begin{pmatrix}1 & 0 \\
-e^{\imath h(k)} & 1 \\
\end{pmatrix}.
\end{align}
and by the same manipulation of the jump contour as before, simply remove that region where the conditions above are met.  

Finally, there may be a region where neither of the conditions above can be made to be met. In this case one may apply the following condition $h'(k)=0$ and $-\log(a_+a_-)<\Re (g_+-g_-)<0. $ In this region one has
\begin{align}
\tilde G\simeq\begin{pmatrix}0 & e^{-\imath h_0} \\
-e^{\imath h_0} & 0 \\
\end{pmatrix}.\label{NontrivialSeg}
\end{align}

The simplification afforded by this Riemann-Hilbert problem of this form is that we have segments, each of which the jump matrix is constant and of a  specific off-diagonal form, while on other segments there is no jump. Such a problem can be solved by making use of Riemann theta functions, as shall be made explicit below.   

Another point to be made is that for $x>0$ and large $k$ the conditions $h'>0,$ $g_+-g_-=0$ can be met trivially since $\log a_-\to0$. For $x<0$ it is much more convenient to apply the transformation $G\to e^{-\sigma_z \log a_+}G e^{-\sigma_z \log a_-},$ $G_+\to G_+ e^{\sigma_z \log a_+}$ $G_-\to G_- e^{-\sigma_z \log a_-}$ to first bring $G$ to the form:\begin{align}
&G=\begin{pmatrix}\frac{1}{a_+a_-} &  \Lambda e^{-\imath kx-\log(a_+)} \\
-\frac{  1}{ \Lambda   }e^{\imath k x+\log(a_+) } & 1 
\end{pmatrix},
\end{align}
whereupon $h'<0,$ $g_+-g_-=0$ can be applied rather than letting  $g_--g_+=\log(a_+a_-).$ Of course both points of view are equivalent, but the latter one is more symmetric with respect to $x\leftrightarrow -x$.    

The fact that  $g_\pm$ can be found is related to the fact that it satisfies a certain scalar Riemann-Hilbert problem. We shall not go over this here, as the method is described in Ref. \cite{Deift:Zhou:Steepest:Descent} in full. We do mention here, that practically $g_\pm$ can constructed in a self consistent manner which will be described presently. The method thus described then actually provides the solution to the inviscid equations, Eqs. (\ref{inviscid1}, \ref{inviscid2}) in the case where the final Riemann-Hilbert problem contains just one non-trivial segment with jump matrix of the form of Eq. (\ref{NontrivialSeg}). Then the endpoints of the segments are $\lambda_1,$ $\lambda_2$ that obey Eqs. (\ref{RiemannInvariant1}, \ref{RiemannInvariant2}), which are equivalent to Eqs. (\ref{inviscid1}, \ref{inviscid2}).

Since the case where only one segment is already very instructive, and since additional segments only encumber the notation, we concentrate on this case. However, it is quite easy, once the case of one segment is understood to generalize the procedure to several segments. We thus use notations in this sub-section which suggest a single segment, with the understanding that the generalization is straightforward. 

\begin{figure}[ht]
\begin{center}
\includegraphics[width=0.60\textwidth,clip=]{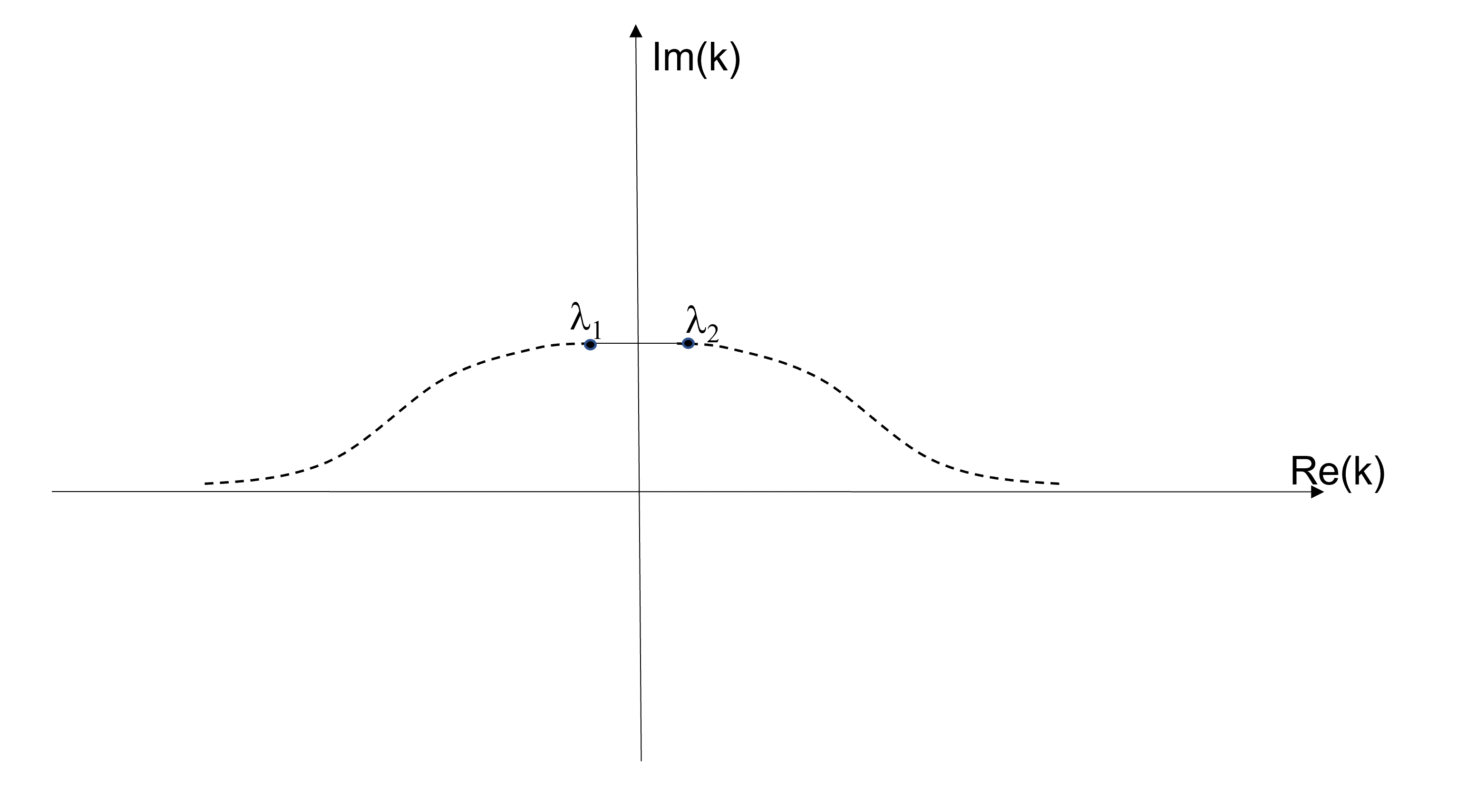}
\end{center}
\caption{The jump contour after deformation. The actual jump is only on the segment $[\lambda_1,\lambda_2]$, the rest of the contour, which is denoted in a dashed line has no jump on it, due to the procedure which removes it by decomposing $G$\ into $G_1G_2$.   }
\label{howcontourlooks}
\end{figure}

It further turns out that, in order to achieve all the above conditions,  the segment $[\lambda_1,\lambda_2]$ does not lie on the real axis. This is already suggested in the transformation to $\rho$ and $v$ in Eq. (\ref{transformToInvariants}), where $\rho$ and $v$\ are real but then $\lambda_i$ is not. In fact the segment is of the form $[\frac{\imath v}{2}-\sqrt{\rho},\frac{\imath v}{2}+\sqrt{\rho}],$ where $v$ and $\rho$ are real, and $\rho\geq0$. This means that the jump contour is also deformed to the complex plane. We give a description of the contour in Fig. \ref{howcontourlooks}. 

Let us denote the endpoints of the segment by $\lambda_1$ and $\lambda_2$. We definte:
\begin{align}
R_2(k)=\sqrt{k-\lambda_1}\sqrt{k-\lambda_2}.
\end{align}
In fact $g_\pm'$ must obeys:
\begin{align}
&g_++g_-=\imath k x-\log(a_-(k))-\log(\Lambda) & k\in[\lambda_1,\lambda_2],\\&g_+-g_-=0 &\mbox{otherwise}.
\end{align}
One may solve these conditions by taking $g_\pm$ to be given by a single function $g$ defined  on the complex with a cut on the segment $[\lambda_1,\lambda_2],$ while $g_\pm$ are just the values of this single function above and below the cut. Such a function may be given by naturally associating with it a differential and writing\begin{align}
& \label{gprimethroughw}g'(k)dk=\frac{dw(k)}{R_2(k)} \end{align}
where $dw(k)$ itself has a jump discontinuity of value $R_2(\mu)\left(\imath x-(t-1)2\mu+\frac{a_\pm'(\mu)}{a_\pm(\mu)}\right)$ over the branch cut at $[\lambda_1,\lambda_2],$ and is smooth everywhere else. Such a differential can be written as:  
\begin{align}
dw(k)=dk
\int_{\lambda_1}^{\lambda_2} \frac{\sqrt{(\mu-\lambda_1)(\mu-\lambda_2)}}{\mu\mp\imath 0^+ -k} \left(\imath x-(t-1)2\mu+\frac{a_\sigma'(\mu)}{a_\sigma(\mu)}\right)\frac{d\mu}{2\pi\imath},
\end{align}
where we have allowed for using either $a_+$ or $a_-$ by making use of $a_\sigma$ where $\sigma=\pm$, respectively. 
or alternatively as:
\begin{align}
dw(k)=R_2(k)\left(\imath x-(t-1)2k+\frac{a_\sigma'(k)}{a_\sigma(k)}\right)dk-dw_{\rm sing}(k)
\end{align}
where $dw_{\rm sing}(k)$, which is adorned with the subscript $\rm sing$, denotes the singular part of terms which precede it, namely all the singularities of those terms away from the branch cut when this term is treated as a differential . These singularities are  the poles of the expression at infinity  
and the branch cut on  $S$. Explicitly we have:
\begin{align}
&\frac{dw_{\rm sing}(k)}{dk} = x(k-\imath v/2)+2\imath (t-1)\left(k(k- \imath v/2)-\frac{\rho}{2}        \right)+\sigma\int_S \frac{m'(\mu) R_2(\mu)}{k-\mu}  \frac{d\mu}{2\pi\imath }
\end{align}

In order for $g'$ defined in Eq. (\ref{gprimethroughw}) not to diverge at the branch points $\lambda_i$, we must demand:
\begin{align}
dw_{\rm sing}\left(\lambda_i\right)=0.
\end{align}
Setting $\lambda_i=\frac{\imath v}{2} \pm \sqrt{\rho}$ in this equation, one gets more explicitly:
\begin{align}
 \pm\sqrt{\rho}\left(x-v  (t-1)\right)+\imath\rho  (t-1)+\sigma\int_S \frac{m'(\mu) R_2(\mu)}{\frac{\imath v}{2}\pm\sqrt{\rho}-\mu}  \frac{d\mu}{2\pi\imath }=0.
\end{align}
Summing the two equations and subtracting them (followed by a division by $\sqrt{\rho}$) leads to the the following:
\begin{align}
&\label{GeneralSaddlePoint1}x-v  (t-1)=\sigma\int_S \frac{m'(\mu) R_2(\mu)}{(\mu-\frac{\imath v}{2})^2-\rho{}}  \frac{d\mu}{2\pi\imath },\\
&\label{GeneralSaddlePoint2}\rho  (t-1)=\imath \sigma\int_S \frac{(\mu-\frac{\imath v}{2})m'(\mu) R_2(\mu)}{(\mu-\frac{\imath v}{2})^2-\rho}  \frac{d\mu}{2\pi\imath }.
\end{align}
Note that the second equation always has the solution $\rho=0$, as the integral on the right hand side vanishes at $\rho=0$ due  the symmetry $\mu\to-\mu^*$, which $m(\mu)$ and the set $S$\ respect, but with respect to which the integral is antisymmetric. The solution of these equations are solutions the inviscid equations, Eqs. (\ref{inviscid1}, \ref{inviscid2}), above as shall be shown below.  

If set $t=1$ and assume $\rho$ to be small, we obtain from these equations the two conditions: 
\begin{align}
&\label{t=1vEq}x=\sigma\int_S \frac{m(\mu)}{(\mu-\imath v/2)^2}  \frac{d\mu}{2\pi\imath },\\
&\label{t=1rhoEq}0= \int_S\frac{m(\mu)}{(\mu-\imath v/2)^3}\frac{d\mu}{2\pi\imath}\mbox { or  } \rho=0.
\end{align}In fact, we know that at $t=1$ we have $\rho=0,$ for any $x\neq0$, Namely, the second condition, given in Eq. (\ref{t=1rhoEq}), may be only solved by $\rho=0$ for $x\neq0.$ To show that this indeed the case, we take a  derivative with respect to $x$ of the first condition, and obtain:
\begin{align}
1=\imath \partial_x v\int_S \frac{m(\mu)}{(\mu-\imath v/2)^3}  \frac{d\mu}{2\pi\imath }
\end{align} 
Namely, if the integral in the second condition, Eq. (\ref{t=1rhoEq}),  is to vanish then  then $\partial_x v$ must diverge. Since this is not the case for any finite $x$ by assumption, we must conclude that $\rho(x,1)=0$ for all $x\neq0$.

More generally and for any $t$ there exists a region in which $\rho=0,$ for this region the Hodograph equations (Eqs.(\ref{GeneralSaddlePoint1}, \ref{GeneralSaddlePoint2})\ above)\ reduce to
\begin{align}
x+\frac{\imath v}{2}  (t-1)=\sigma\int_S \frac{m(\mu)}{(\mu-\imath v/2)^2}  \frac{d\mu}{2\pi\imath }.\label{simplehodo}
\end{align} 
It is often the case that one may solve the large $\kappa $ (or, equivalently, large $\Lambda$) limit at $t=1$, where $\rho=0$ for $x\neq0$ and obtain that the left hand side of Eq. (\ref{simplehodo}) is equal to a given function of $v$. Then it is a matter of solving a singular integral equation in order to find $m(\mu)$. This $m(\mu)$\ plays also a role in an exact solution of the problem, where it is defined by Eq. (\ref{mExactDef}). Of course the exact $m(\mu)$ may have small corrections which are more difficult to obtain by examining the approximate solution (although it is in principle possible to develop a systematic expansion), nonetheless, the approximate $m(\mu)$\ may either prove exact, or may supply a valuable first guess to find an exact solution. We demonstrate this in section \ref{ExampleSection} below.

Eqs. (\ref{GeneralSaddlePoint1},\ref{GeneralSaddlePoint2}) are general equations for $v$ and $\rho$, or equivalently, for $\lambda_1$ and $\lambda_2$. We  wish to show  now that these equations are solutions to Eqs. (\ref{inviscid1}, \ref{inviscid2}) or, equivalently,  to Eqs. (\ref{RiemannInvariant1}, \ref{RiemannInvariant2}). 

The requirement that $g'$ does not diverge suggests that around $\lambda_i$ it has the form $g'(k)\sim \sqrt{k-\lambda_i}.$ Let us define $d\Omega_x$ and $d\Omega_t$ as follows:  
\begin{align}
\frac{\partial g'}{\partial x}dk=\imath d\Omega_x, \quad \frac{\partial g'}{\partial t}dk=-2 d\Omega_t.
\end{align}
The compatibility condition for these two equations read:
\begin{align}
\imath \partial_t d\Omega_x=2\imath \partial_x d\Omega_t.\label{Compatibility}
\end{align}
Now $d\Omega_x$ and $d\Omega_t$ can be easily seen to be meromorphic differential on the genus-0 Riemann surface associated with $R_2(k)$ with pole of order $2$ and $3$\ respectively, and residue $\pm1$  at infinity at the upper and lower sheets of the Riemann surface respectively, which, by uniqueness of such differentials,  means
that we may immediately write:\begin{align}
d\Omega_x=\frac{k-\frac{\lambda_1+\lambda_2}{2}}{\sqrt{k-\lambda_1}\sqrt{k-\lambda_2}} dk, \quad d\Omega_t=\frac{k\left(k-\frac{\lambda_1+\lambda_2}{2}\right)-\frac{(\lambda_2-\lambda_1)^2}{2}}{\sqrt{k-\lambda_1}\sqrt{k-\lambda_2}} dk.\label{dOmegaxt} 
\end{align}
Indeed expanding at infinity we get $d\Omega_x\sim \left(\pm1+O(1/k^2)\right)dk$ and  $d\Omega_t\sim \left(\pm k +O(1/k^2)\right)dk$ on the upper and lower sheet, respectively.

If the explicit expressions for $d\Omega_x$ and $d\Omega_t$ are substituted into the compatibility equation, Eq. (\ref{dOmegaxt}), one obtains equations for $\lambda_i$ by examining the behavior around $\lambda_i$ of both sides of the compatibility condition, Eq. (\ref{Compatibility}). Indeed, one obtains a term that diverges as $(k-\lambda_i)^{-3/2}$ around $\lambda_i$ and the residue of that divergence must coincide on both sides of the equation. This condition then reads:
\begin{align}
\left(\lambda_i-\frac{\lambda_1+\lambda_2}{2}\right)\partial_t\lambda_i=2\imath\left(\lambda_i\left(\lambda_i-\frac{\lambda_1+\lambda_2}{2}\right)-\frac{(\lambda_2-\lambda_1)^2}{8}\right)\partial_x\lambda_i
\end{align}
Dividing both sides by $\lambda_i-\frac{\lambda_1+\lambda_2}{2}$ yields immediately the inviscid equation encountered above for the Riemann invariants $\lambda_1, $ $\lambda_2$ of Eqs. (\ref{inviscid1}, \ref{inviscid2}), namely equations (\ref{RiemannInvariant1}, \ref{RiemannInvariant2}). These equations read as follows: 
\begin{align}
&\partial_t \lambda_i=\left(\frac{3}{2}\lambda_i+\frac{1}{2}\lambda_{3-i}\right)\imath \partial_x\lambda_i.
\end{align}

Note that, although we have concentrated on the case where only two Riemann invariants are present, the same procedure will yield equations with any (even) number of Riemann invariants, which generalize  equations (\ref{RiemannInvariant1}, \ref{RiemannInvariant2}), to the case where the reduced Riemann-Hilbert contour is the union $\cup_{j=1}^{g+1}[\lambda_{2j-1},\lambda_{2j}]$, where $g$ is the genus of the Riemann surface associated with $R_{2g+2}(k)\equiv\prod_{j=1}^{g+1}\sqrt{(k-\lambda_{2j-1})(k-\lambda_{2j})} $. The $\lambda_i$'s thus obtained are then moduli of oscillatory solutions of the original nonlinear Schr\"odinger equation, Eqs. (\ref{NLS1}, \ref{NLS2}).  These oscillatory solutions are given in Eq. (\ref{qNLSoscill},\ref{pNLSoscill}) below. When two of the $\lambda_i$'s coincide, namely when $\lambda_{2j-1}\to\lambda_{2j},$such an oscillatory solution takes on a solitonic nature. The only added component here is that the differentials $d\Omega_x$ and $d\Omega_t$ may be shown to be normalized as to have null $a-$cycles.   

\subsection{Solution of the Reduced Problem}

The solution of the reduced Riemann-Hilbert problem is achieved by making use of the Riemann theta functions. One can find more details about this solution in Refs \cite{Deift:Zhou:Steepest:Descent,belokolos:Bobenko:Algebro:Geometrical:Integrable}, while here we give  merely  a very rapid exposition.  It should be noted that this section is given here only for completeness, since if one is only interested in the case of two $\lambda_i$'s, then the identification of the Riemann invariants as the end-points of the reduced Riemann-Hilbert contour has already been made, albeit without justification (it was merely suggestive that the two objects, the Riemann invariants of Eqs.(\ref{inviscid1}, \ref{inviscid2}) and the endpoints of the reduced Riemann-Hilbert contour,  obey the same differential equations, Eqs. (\ref{RiemannInvariant1}, \ref{RiemannInvariant2})), and, furthermore, the solution of the Riemann-Hilbert problem is from this point on standard\cite{Deift:Zhou:Steepest:Descent,belokolos:Bobenko:Algebro:Geometrical:Integrable}, namely,  there are no special features associated with the peculiar formulation of the large deviation problem, except the fact that the Riemann-Hilbert contour lies rather unconventionally away from the real axis, the reflection across the imaginary axis being the symmetry that is obeyed by the contour in this case. This situation due to the non-conventional real section of the non-linear Schr\"odinger equation afforder by two real fields, $p$ and $q,$ rather than by two fields which are complex conjugates of each other.   

 In order to introduce the solution assume that the Riemann surface at hand is given by
\begin{align}
R_{2g+2}(k)=\prod_{i=1}^{2(g+1)}\sqrt{k-\lambda_i}.
\end{align}
In this case we have $g$ holomorphic differentials $\omega_i$ which may be normalized as follows:
\begin{align}
\oint_{a_j} \omega_i=2\pi\imath \delta_{ij}, 
\end{align}
\begin{figure}
\begin{center}
\includegraphics[width=0.80\textwidth,clip=]{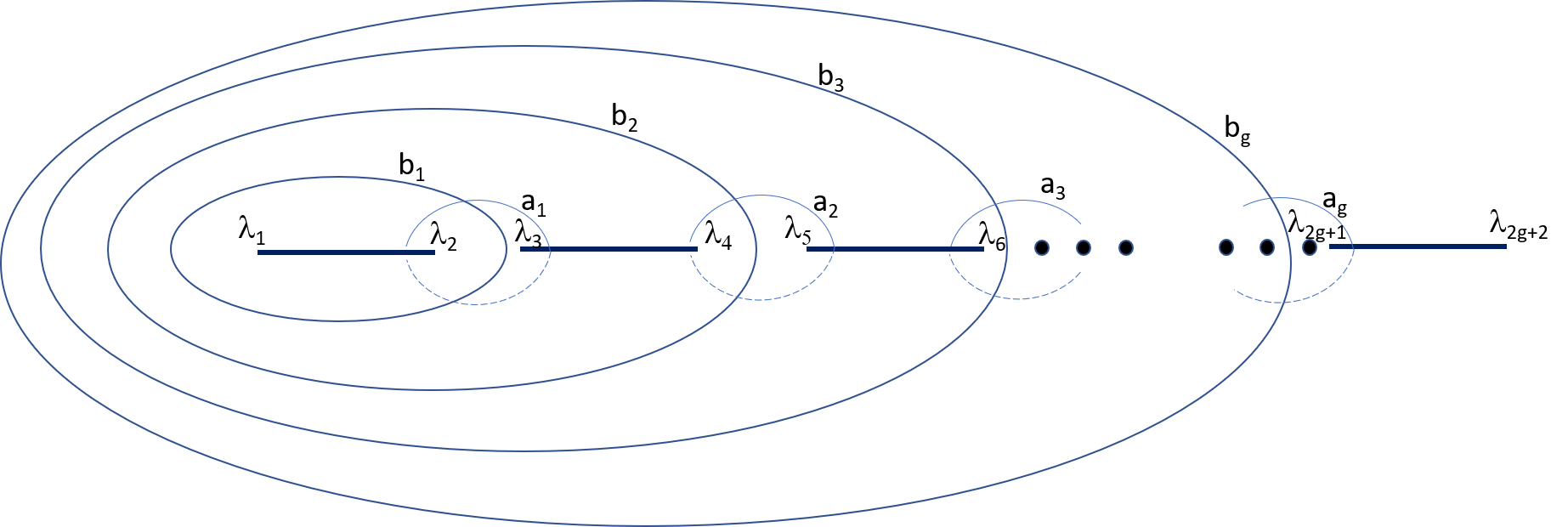}
\end{center}
\caption{The cycles over the Riemann surface associated with the function $R_{2g+2}(k)$. Dashed lines describe the part of the cycle that lies on the lower sheet.    }
\label{cycles}
\end{figure} 
where $a_i$ denotes an $a-$ cycle conventionally defined as shown in Fig. \ref{cycles}. 

Then one defines the Abel map:
\begin{align}
A_i(P)=\int_{\infty_+}^P \omega_i.
\end{align}
Namely the integral is taken from $\infty$ on the upper sheet to the point $k$. Thus $\bm A(k^\pm)$ denotes that the point $P$\ is to be taken as the point $k$ on the upper or lower sheets respectively. One further defines the Riemann matrix 
\begin{align}
B_{ij}=\oint_{b_j} \omega_i
\end{align}
The Riemann theta function is defined as\cite{belokolos:Bobenko:Algebro:Geometrical:Integrable}:
\begin{align}
\theta(\bm A)=\sum_{\bm m\in \mathds{Z}^g} \exp\left[\frac{1}{2} \bm m \bm B\bm m+\bm m \bm A \right].
\end{align}

We also define $\bm A_g$ to be a vector whose $i-$th element is given by $\oint _{b_i} g'dk. $ Lastly, we define a differential $d\Omega_N$ to be a meromorphic differential with pole at infinity on the upper and lower sheets and of residue $\pm1,$ respectively.  Associated with this definition is the vector $\bm V$ a vector whose $i-$th element is given by  $\oint _{b_i} d\Omega_N.$ In the lore of algebraic Riemann surfaces, the following  is a standard identity:

\begin{align}
\bm V=\oint_{\bm b} d \Omega_N=\int_{\infty_+}^{\infty_-}d\bm\omega=\bm A(\infty_-).
\end{align}  

Is convenient to find a solution to the matrix $M_\pm$ instead of $G_\pm$ , where $M_\pm$ is defined below: \begin{align}
M_\pm= G_\pm\begin{pmatrix}k & 0 \\
0 & 1 \\
\end{pmatrix},
\end{align}
where  this change of variable results in the following asymptotics as $k\to\infty$ : $M_\pm\to\begin{pmatrix}k & k^{-1} \\
1 & 1 \\
\end{pmatrix}.$ But otherwise the matrix $M_\pm$ solves the same Riemann-Hilbert problem as $G_\pm$, which becomes the reduced Riemann-Hilbert problem in the limit of large $\Lambda$.

The solution reads:
\begin{align}
M_\pm=\begin{pmatrix}\frac{\theta(\bm D-\bm V)\theta(\bm A_g+\bm A(k^+)-\bm V+\bm D)}{\theta(\bm A(k^+)-\bm V+\bm D)\theta(\bm A_g-\bm V+\bm D)} e^{g(k)-\int^k \Omega_N} &\frac{\theta(\bm D-\bm V)\theta(\bm A_g+\bm A(k^-)-\bm V+\bm D)}{\theta(\bm A(k^-)-\bm V+\bm D)\theta(\bm A_g-\bm V+\bm D)} e^{-g(k)+\int^k \Omega_N}  \\
\frac{\theta(\bm D)\theta(\bm A_g+\bm A(k^+)+\bm D)}{\theta(\bm A(k^+)+\bm D)\theta(\bm A_g+\bm D)} e^{g(k)} & \frac{\theta(\bm D)\theta(\bm A_g+\bm A(k^-)+\bm D)}{\theta(\bm A(k^-)+\bm D)\theta(\bm A_g+\bm D)}e^{-g(k)} \\
\end{pmatrix},
\end{align}
here $\bm D$ is a constant vector depending neither on time, space or auxiliary spectral parameter, $k$.

The asymptote is therefore:
\begin{align}
\label{MplusExpansion}M_\pm=&\left(\mathds 1+k^{-1}\Psi_1+\dots\right)\begin{pmatrix} 0 & e^{\imath xk-tk^2} \\
k e^{\imath xk-tk^2} & 0 \\
\end{pmatrix}\end{align}where \begin{align}\label{Psi1}\Psi_1=\begin{pmatrix}\dots & \frac{\rho_0\theta(\bm A_g-\bm V+\bm D)\theta(\bm D)e^{ -\int \imath k_0dx-\omega_0 dt}}{\theta(\bm D-\bm V)\theta(\bm A_g+\bm D)} \\
\frac{\theta(\bm D)\theta(\bm A_g+\bm V+\bm D)e^{ \int \imath k_0dx-\omega_0 dt}}{\theta(\bm V+\bm D)\theta(\bm A_g+\bm D)} & \dots \\
\end{pmatrix},
\end{align} 
and the ellipsis denotes terms unimportant for the sequel. This expression makes use of the following definitions of $v_0$, $\omega_0$ and $\rho_0$, connected to the asymptotes of $\Omega_i(k)\equiv \int_{\lambda_1}^k d\Omega_i,$ with $i\in\{x,t,N\}$
\begin{align}
&\label{LArgeOmegaXP} \Omega_x(z)=\pm\left(z-\imath \frac{v_0}{2} \right)+O(1/z), \quad \Omega_t(z)=\pm(2z^2-\omega_0)+O(1/z),
\\& 
\label{LargeOmegaN}\Omega _N(z)=\pm\log\frac{2z}{\sqrt{\rho_0}} +O(1/z)\end{align}
Substituting Eqs. (\ref{MplusExpansion}, \ref{Psi1}) into Eq. (\ref{Gpmxderiv}), one may deduce:
\begin{align}
\Psi_1=\begin{pmatrix}\dots &- \frac{p}{8}    \\
 q & \dots \\
\end{pmatrix}
\end{align}Thus we have:
\begin{align}
&\label{qNLSoscill}q=\frac{\theta(\bm D)\theta(\bm A_g+\bm V+\bm D)e^{ \int \imath k_0dx-\omega_0 dt}}{\theta(\bm V+\bm D)\theta(\bm A_g+\bm D)}\\
&\label{pNLSoscill}p=-2\rho_0\frac{\theta(\bm A_g-\bm V+\bm D)\theta(\bm D)e^{ -\int \imath k_0dx-\omega_0 dt}}{\theta(\bm D-\bm V)\theta(\bm A_g+\bm D)}
\end{align}

In the genus $0$ case ($g=0$), the theta function degenerates to a constant. And computing the asymptotes in Eqs. (\ref{LArgeOmegaXP},\ref{LArgeOmegaXP}) explicitly gives $ v_0=v,$ $\rho_0=\rho$ and $\omega_0=2\rho-\frac{v^2}{4}$ which recovers  the Cole-Hopf transformation in Eq. (\ref{ColeHopf}) $q= e^{-\frac{1}{2}\int^x v},$ $p=-2 \rho e^{\frac{1}{2}\int^x v}$ .

\section{Example for Flat Initial Conditions\label{ExampleSection}}In Ref. \cite{Meerson:Katzav:Vilenkin} the boundary value problem:
\begin{align}
\rho(x,1)=\frac{16\kappa^3}{3\pi}  \delta(x), \quad v(x,0)=0, 
\end{align}
was considered in the inviscid limit. This is term "flat" initial conditions since $v$ is constant at initial time.  In Ref. \cite{Meerson:Katzav:Vilenkin} Eqs. (\ref{inviscid1},\ref{inviscid2}) were solved with these boundary conditions. The solution may be written by first solving $\beta(t)$ from the equation:
\begin{align}
t=\frac{2}{\pi\beta(t)}\sqrt{\beta(t)-1}+\frac{2}{\pi} \arctan\sqrt{\beta(t)-1}.\end{align}Then $\rho$ and $v$ can be written explicitly as: \begin{align}
&\rho(x,t)=\begin{cases}\beta\left(\kappa^2-\frac{\pi^2\beta ^2}{16}x^2\right) & |x|<\frac{4\kappa}{\pi\beta} \\
0  & |x|>\frac{4\kappa}{\pi\beta} \\
\end{cases}\\
&v(x,t)=\begin{cases}- \frac{\pi}{2}\beta\sqrt{\beta-1}x & |x|<\frac{4\kappa}{\pi\beta} \\
\mbox{solves }x-vt=- \frac{2}{\pi}\left(2\kappa +v \arctan\frac{v}{2\kappa}\right)\sigma_v & \frac{4\kappa}{\pi    }>|x|>\frac{4\kappa}{\pi\beta} \\
0& \frac{4\kappa}{\pi    }<|x|
\end{cases}.
\end{align}

        To identify the asymptotic Riemann-Hilbert problem associated with this large $\Lambda$ limit, we may take the solution for $v(x,1)$  identify it as Eq. (\ref{simplehodo}), with $\sigma_v=-\sigma$ and  $k=\frac{\imath v}{2}$. Then one has:
\begin{align}
\int^\kappa_{-\kappa} \frac{m'(\mu)}{\mu-\frac{\imath v}{2}}\frac{d\mu}{2\pi }=v-\frac{2}{\pi}\left(2\kappa +v \arctan\frac{v}{2\kappa}\right).
\end{align}
This is a singular integral equation which can be solved by standard means, we, however, may guess the solution, namely $m'(\mu)=-4\mu,$ which suggests $ m(\mu)=2(\kappa^2-\mu^2)$.  
This coincides with the result of Ref. \cite{Krajenbrink:Doussal:Flat:Brownian} where it was shown that within an exact solution one obtains $m(\mu)=2(\kappa^2-\mu^2)-2\log(\mu)$, the last term being a logarithmic correction to out approximate solution where the large parameter is $\mu\sim \kappa.$ 

        Having identified $m(\mu)$, we may now check to see if Eqs. (\ref{GeneralSaddlePoint1},\ref{GeneralSaddlePoint2}), which may be considered as the result of integration of the inviscid differential  equations,  are satisfied in this case. This is a matter of substituting $m'(\mu)=-4\mu$ in those equations and performing the integrals. One find the following:
\begin{align}
&x-v\left( t+\frac{2}{\pi} \arg \frac{{}\sqrt{\kappa-\lambda_2}-\sqrt{\kappa-\lambda_1}}{\sqrt{\kappa-\lambda_1}+{}\sqrt{\kappa-\lambda_2}}\right)-\frac{4}{\pi} \Re\left[ R_2(\kappa)\right]=0\\&\pi\rho  \left( t+\frac{2}{\pi} \arg \frac{{}\sqrt{\kappa-\lambda_2}-\sqrt{\kappa-\lambda_1}}{\sqrt{\kappa-\lambda_1}+{}\sqrt{\kappa-\lambda_2}}\right)- \Im\left[ \left(2\kappa+\imath v\right)R_2(\kappa)\right]=0, \end{align}
and the fact that these equations are indeed satisfied can be checked by direct substitution.

\section{Conclusion}
In this paper we have established a connection between the inverse scattering method and the Whitham limit in the case of the boundary value problems that appear in certain large deviation problems. We have tried to give all the important features of the approach that connects the two problems, while leaving out many of the specific details that apply to certain cases. First, we have only dealt with the case where the relevant nonlinear equations to be solved are the non-linear Schr\"odinger equations, with  real fields. We believe that the generalization to such systems as the derivative non-linear Schr\"odinger equations is not substantially different than the current case.

 Furthermore we have mainly dealt with the case where the strongly non-linear limit leads to inviscid equations, where dispersive terms may simply be dropped. Although at first sight it may seem that this case is rather special, as it is known that instabilities, such as shocks, can cause oscillations or solitons to appear in the solution, it is actually quite straightforward to generalize the method to such cases, since the form of the solution in the case where the Riemann-Hilbert contour is multi-segmented is written down in Eqs. (\ref{qNLSoscill},\ref{pNLSoscill}). The case of solitons appear when two enpoints of the Riemann-Hilbert contour meet, as is well known. Indeed, in that limit the theta functions appearing in  Eqs. (\ref{qNLSoscill},\ref{pNLSoscill}) degenerate into hyperbolic trigonometric functions, from which solitons are easily obtained. In the case of multi-segmented Riemann-Hilbert contours one can recover a generalization of the equations for the Riemann invariants, Eqs. (\ref{RiemannInvariant1}, \ref{RiemannInvariant2}), by making use Eqs. (\ref{Compatibility}), by following the procedure outlined in this paper. Such a procedure is well known from Refrs. \cite{Gurevich:Pitavsk,88:Flaschka:KdV:Avging,Flaschka:McLaughlin:Canonical:Conjugate:Variables,83:Krichever:Averaging} .

\section{Acknowledgement}
I\ wish to thank Baruch Meerson  for many useful discussions. I\ wish to acknowledge the  Binational Science Foundation which has supported this research through grant number 2020193. 

\end{document}